\documentclass[10pt]{article}
\usepackage{amsmath}
\usepackage{amsfonts}
\usepackage{authblk}
\usepackage{bm}
\usepackage{booktabs}
\usepackage{braket}
\usepackage[labelfont=bf]{caption}
\usepackage[inline]{enumitem}
\usepackage{float}
\usepackage[a4paper, total={6.5in, 9in}]{geometry}
\usepackage{graphicx}
\usepackage{mdframed}
\usepackage[numbers,sort]{natbib}
\definecolor{SLACRed}{RGB}{139,24,27}
\usepackage{hyperref}
\hypersetup{colorlinks,breaklinks,allcolors = SLACRed}
\usepackage{xcolor}
\usepackage{stmaryrd}
\usepackage[normalem]{ulem}

\newcommand{\ud}{\mathrm{d}}
\newcommand{\trans}{\scriptscriptstyle\top}

\newcommand{\meV}{\mskip3mu\mathrm{meV}}
\newcommand{\ps}{\mskip3mu\mathrm{ps}}
\DeclareMathOperator*{\argmax}{arg\,max}

\title{Machine learning enabled experimental design and parameter estimation for ultrafast spin dynamics}
\author[1,2]{Zhantao Chen}
\author[1]{Cheng Peng}
\author[1,2]{Alexander N.\@ Petsch}
\author[1,3]{Sathya R. Chitturi}
\author[4]{Alana Okullo}
\author[4]{Sugata Chowdhury}
\author[2]{Chun Hong Yoon}
\author[1,2,*]{Joshua J.\@ Turner}
\affil[1]{Stanford Institute for Materials and Energy Sciences, Stanford University, Stanford, CA, USA.}
\affil[2]{Linac Coherent Light Source, SLAC National Accelerator Laboratory, Menlo Park, CA, USA.}
\affil[3]{Department of Materials Science and Engineering, Stanford University, Stanford, CA, USA.}
\affil[4]{Department of Physics and Astronomy, Howard University, Washington DC, USA.}
\affil[*]{Corresponding author: \href{mailto:joshuat@slac.stanford.edu}{\textnormal{\texttt{joshuat@slac.stanford.edu}}}}

\begin{document}
\maketitle
\begin{abstract}
Advanced experimental measurements are crucial for driving theoretical developments and unveiling novel phenomena in condensed matter and material physics, which often suffer from the scarcity of facility resources and increasing complexities. To address the limitations, we introduce a methodology that combines machine learning with Bayesian optimal experimental design (BOED), exemplified with x-ray photon fluctuation spectroscopy (XPFS) measurements for spin fluctuations. Our method employs a neural network model for large-scale spin dynamics simulations for precise distribution and utility calculations in BOED. The capability of automatic differentiation from the neural network model is further leveraged for more robust and accurate parameter estimation. Our numerical benchmarks demonstrate the superior performance of our method in guiding XPFS experiments, predicting model parameters, and yielding more informative measurements within limited experimental time. Although focusing on XPFS and spin fluctuations, our method can be adapted to other experiments, facilitating more efficient data collection and accelerating scientific discoveries.
\end{abstract}

\section{Introduction}

Ever since the discovery of x-rays, considerable breakthroughs have been made using them as a probe of matter, from testing models of the atom to solving the structure of deoxyribonucleic acid (DNA). Over the last few decades with the proliferation of synchrotron x-ray sources around the world, the application to many scientific fields has progressed tremendously and allowed studies of complicated structures and phenomena like protein dynamics and crystallography \cite{hendrickson2000synchrotron, martin2021protein}, electronic structures of strongly correlated materials \cite{fadley2010x, neppl2015time}, and a wide variety of elementary excitations \cite{burkel2000phonon, ament2011resonant}. With the the development of the next generation of light sources, especially the x-ray free electron lasers (X-FEL) \cite{ribic2012status, pellegrini2017x}, not only have discoveries accelerated, but completely novel techniques have been developed and new fields of science have emerged, such as laboratory astrophysics \cite{takabe2021recent,hwang2021x,husband2021x,wark2022femtosecond} and single particle diffractive imaging \cite{bogan2008single,donatelli2017reconstruction, bielecki2020perspectives}. 

Among these emerging techniques brought by X-FELs, the development of x-ray photon fluctuation spectroscopy (XPFS) holds particular relevance for condensed matter and material physics \cite{shen2021snapshot}. XPFS is a unique and powerful approach that opens up numerous opportunities to probe ultrafast dynamics of timescales corresponding to the $\mu$eV to meV-energy level. As the high-level coherence of the x-ray beam encodes subtle changes in the system at these timescales, XPFS is capable of investigating fluctuations of elementary excitations, such as that of the spin \cite{plumley2023ultrafast}. The fluctuation spectra collected using this method can be directly related back to correlation functions derived from Hamiltonians \cite{lehmkuhler2021femtoseconds,mohanty2022computational}, yielding invaluable experimental insights for theoretical developments and deeper understandings of the underlying physics.

Despite the breakthroughs, the critical dependence of XPFS on the rare experimental resource of X-FEL beamtime has prevented widespread adoption of such advanced XPFS measurements and hindered further scientific explorations. The targeting measurements of fluctuating dynamics are often complicated. For instance, in the study of ultrafast spin fluctuations, a multitude of excitation modes can exist simultaneously, leading to complex time-dependent signals that requires many delay-time measurements to accurately capture the oscillatory and decaying profiles that reveal the crucial physical information and quantitatively inform model parameters. The scarcity of beamtime resources and complicated measurement signals underscore the vital importance of theory-informed and data-driven experimental design methods when utilizing techniques such as XPFS for the study of ultrafast fluctuations.

One such method to aid in the collection and interpretation of data such as that generated in XPFS is the Bayesian optimal experimental design (BOED), a type of well-established statistical methods \cite{granade2012robust, huan2013simulation, ryan2016review, mcmichael2022simplified} whose significant potential for real experimental applications has only been harnessed recently \cite{dushenko2020sequential, mcmichael2021sequential, caouette2022robust}. To achieve the most informed experimental design, physically realistic forward computations of system dynamics directly from model Hamiltonian are critically important. However, such forward model evaluations are often computationally intensive and practical applications of BOED can become prohibitive due to the considerable number of forward model computations required for its distribution updates and utility function calculations \cite{long2013fast, ryan2016review, fiderer2021neural}. Therefore, the ability to perform fast and cost-efficient forward model computations is a key factor in the successful incorporation of BOED in XPFS measurements for studying ultrafast dynamics. It is also important to note that the conventional sequential Bayes update method could fail given poorly initialized prior distributions, which are typically based on human input \cite{robert2007prior, gelman2013bayesian}. As such, the implementation of a distribution correction mechanism is also crucial to the successful application of BOED besides the rapid forward computation, ensuring that errors in the initial parameter estimation can be accurately identified and rectified during the data collection.

Some recent progress has hinted at the potential for achieving more computationally efficient BOED by incorporating machine learning techniques into the workflow of Bayesian experimental designs and parameter estimations \cite{fiderer2021neural, nolan2021machine, cimini2023deep}. However, the combined ML-BOED methods tailored for XPFS and ultrafast dynamics are still waiting to be developed. In this work, we introduce a machine learning (ML)-enabled BOED approach, specifically designed to guide measurements of ultrafast dynamics with XPFS and one which will greatly drive advances in this field. For concreteness, we focus our interests on ultrafast spin fluctuations for a realistic Hamiltonian, modeled for XPFS. Our method relies on the use of a neural network as an efficient and accurate surrogate model for linear spin wave theory (LSWT). This model facilitates precise evaluations of distributions and utility function calculations and thereby enables LSWT-guided, real-time Bayesian design and estimation. Furthermore, with the automatic differentiable forward model, gradient descent (GD)-based parameter estimations become feasible and can be naturally incorporated into BOED to achieve distribution corrections and more robust parameter estimations under various experimental conditions. We demonstrate the performance of our method through a comprehensive benchmarking using simulated experimental data. Consequently, our approach provides a powerful tool that leverages both physical models and Bayesian analysis for real-time guidance in ultrafast spin fluctuation studies. 

\section{Problem Formulation and Methods} \label{sec:problem_formulation_and_methods}

\begin{figure}[h]
    \centering
    \includegraphics[width=1.0\linewidth]{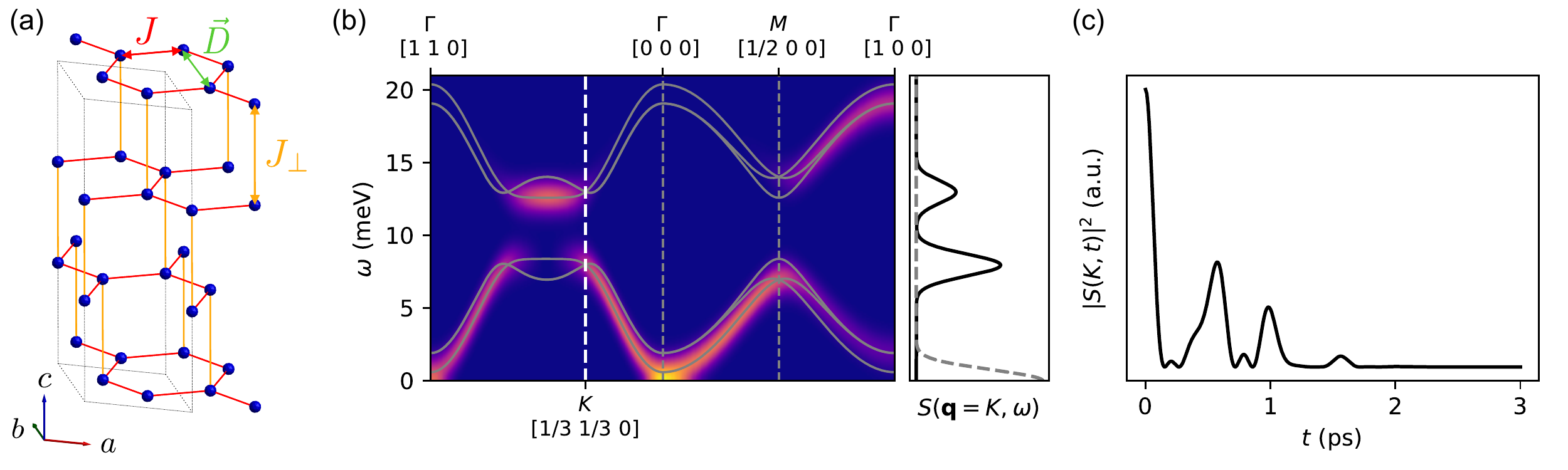}
    \caption{\textbf{Illustration of the physical problem setup.} (a) Van der Waals-layered hexagonal lattice structure and spin interactions (the $z$-axis anisotropy is not shown for clarity). (b) Magnon dispersion $\omega(\mathbf{q})$ and dynamical structure factor $S(\mathbf{q},\omega)$ along a high-symmetry path (left) and its slice at the $K$-point (right). In the right panel, we add the Gaussian-shaped peak (dashed gray curve) to simulate magnon-unrelated intensities captured by detectors. (c) The absolute squared intermediate scattering function (ISF), $|S(\mathbf{q},t)|^{2}$, calculated from both magnon-related and magnon-unrelated scattering signals at the slice of $\mathbf{q}=K$.}
    \label{fig:physical_model}
\end{figure}

In the work described here, we demonstrate our approach by choosing a specific spin model Hamiltonian relevant for van der Waals (vdW) and other 2D magnets. This model contains a relatively complex parameter phase space in the spin interaction degree of freedom and uses an in-plane honeycomb lattice, as illustrated in Figure \ref{fig:physical_model}(a). The spin Hamiltonian is
\begin{equation}
    \mathcal{H}=\sum_{\langle i,j\rangle}\left[J\ \textbf{S}_{i}\cdot \textbf{S}_{j} + \vec{D} \cdot (\textbf{S}_{i}\times \textbf{S}_{j})\right]+\sum_{\langle i,j\rangle_{\perp}}J_{\perp}\ \textbf{S}_{i}\cdot \textbf{S}_{j}+\sum_{j}D_{z}(\textbf{S}_{j}^{z})^{2}
\end{equation}
where the $J$ and $\vec{D}=[0,0,D]^{\trans}$ represent the exchange and Dzyaloshinskii–Moriya (DM) interaction strengths between the nearest neighbors and second nearest neighbors, respectively. The $J_{\perp}$ characterizes the inter-layer exchange coupling and $D_{z}$ defines an easy-axis anisotropy for each spin. This model has been employed to describe inelastic neutron scattering observations of topological spin excitations in materials such as CrI\textsubscript{3} \cite{chen2018topological} and CrXTe\textsubscript{3} (X=Si, Ge) \cite{zhu2021topological}, where the DM interaction is considered responsible for the spin gaps at the Dirac points of the magnon dispersion \cite{owerre2016first,kim2016realization}.

We calculate the spin excitations of Cr\textsuperscript{3+} with spin $S=3/2$ ions on the honeycomb lattice, as shown in Figure \ref{fig:physical_model}(a), and their dynamical structure factors $S(\mathbf{q},\omega)$ from the Hamiltonian in LSWT approximation with the \texttt{SpinW} package \cite{toth2015linear}, since for large spin, the magnetic excitation spectrum is reasonably captured by LSWT. An exemplary calculated magnon dispersion and $S(\mathbf{q},\omega)$ is shown on the left side of Figure \ref{fig:physical_model}(b).

In addition to the magnon-related inelastic peaks in $S(\mathbf{q},\omega)$, we also include a perturbing peak near $\omega=0$ to the ISF calculation as the collected intensities in a real experiment will most certainly include those unwanted elastic or quasi-elastic scattering contributions such as in structural or diffuse scattering signals, which is modeled by a Gaussian-shaped peak for concreteness, as illustrated by the dashed curve in the right panel of Figure \ref{fig:physical_model}(b). More details on this perturbing factor will be presented in Section \ref{sec:results_discussion}. To simulate time-dependent signals that are expected to be extractable from the intensity-intensity correlations determined by XPFS, we use the Fourier-cosine transformation to convert the dynamical structure factor $S(\mathbf{q},\omega)$ to the corresponding ISF $S(\mathbf{q},t)$. For instance, the slice of $S(\mathbf{q},\omega)$ at the $K$-point, indicated by the white dashed line in Figure \ref{fig:physical_model}(b), and its ISF are shown in the right panel of Figure \ref{fig:physical_model}(b) and Figure \ref{fig:physical_model}(c), respectively. 

More specifically, the direct derivative in an XPFS experimental measurement is the contrast function $C(\mathbf{q},t)$, which can be extracted by analyzing photon statistics from the sum of the two intensities $\mathcal{I}(\mathbf{q},\mathtt{t})$ of a pair of successive x-ray probe pulses arriving at $\mathtt{t}=\tau$ and $\mathtt{t}=\tau+t$ \cite{bandyopadhyay-2005-rsi, gutt2009measuring, shen2021snapshot}, i.e.,
\begin{equation}\label{eqn:c2_intensity}
    C(\mathbf{q},t) = \frac{\langle \mathcal{I}^{2}(\mathbf{q},t) \rangle - \langle\mathcal{I}(\mathbf{q},t)\rangle^{2}}{\langle\mathcal{I}(\mathbf{q},t)\rangle^{2}},\qquad \mathcal{I}(\mathbf{q},t)=I(\mathbf{q},\tau) + I(\mathbf{q},\tau+t),
\end{equation}
where $\langle\ldots\rangle$ represents the operation of averaging over different $\tau$, which is obtained by repeating such measurement, i.e., keeping delay time $t$ constant and varying the $\tau$ for each image pair. Meanwhile, the contrast function $C(\mathbf{q},t)$ can be related back to the intermediate scattering function (ISF) $S(\mathbf{q},t)$ through 
\begin{equation}\label{eqn:c2_siegert_like}
    C(\mathbf{q},t) = \beta^{2}\left( \frac{r^{2}+1+2r|S(\mathbf{q},t)|^{2}}{r^{2}+1+2r} \right),
\end{equation}
where $\beta$ is the partial coherence and $r$ represents the intensity ratio of successive pulses \cite{gutt2009measuring}, both are experiment-dependent and are assumed to be known here. By combining Eqs.\@ \eqref{eqn:c2_intensity} and \eqref{eqn:c2_siegert_like}, the fundamental information in $S(\mathbf{q},t)$, or equivalently, $S(\mathbf{q},\omega)$, obtained from XPFS can then be directly compared to theoretical modeling from first principles. It should be noted that the fundamental information obtained from XPFS is identical to that obtained from traditional x-ray photon correlation spectroscopy (XPCS) studies \cite{Gutt-2009-OptExp}; however, XPFS corresponds to a much faster timescale that challenges realistic detector readout and requires a different, and arguably more difficult, experimental data analysis.

We aim to develop a data-driven experiment steering framework that can simultaneously 
\begin{enumerate*}[label=(\roman*)]
    \item estimate Hamiltonian parameters based on current measurements, and 
    \item suggest the next measurement point, time delay $t$, that maximizes the ``information gain''.
\end{enumerate*}
The second point can be roughly understood as the potential to determine the Hamiltonian parameters with as few measurements as possible, and will become clear in Section \ref{sec:boed}. Achieving these two goals requires rapid forward model calculations from the parameter space $x=[J,D]$ to the measurable space $|S(\mathbf{q},t)|^{2}$ conjoined with uncertainty quantification within both spaces, as well as the ability to estimate parameter distributions based on obtained measurements. Therefore, we utilize machine learning to build surrogate models for rapid forward calculations and combine this with recent progresses in Bayesian optimal design algorithms for Hamiltonian parameter estimation and simultaneous experimental decision making \cite{granade2012robust,huan2013simulation,ryan2016review,mcmichael2022simplified}.

\subsection{Machine learning surrogate model for spin excitations} \label{sec:ml_model}

\begin{figure}[h]
    \centering
    \includegraphics[width=0.8\linewidth]{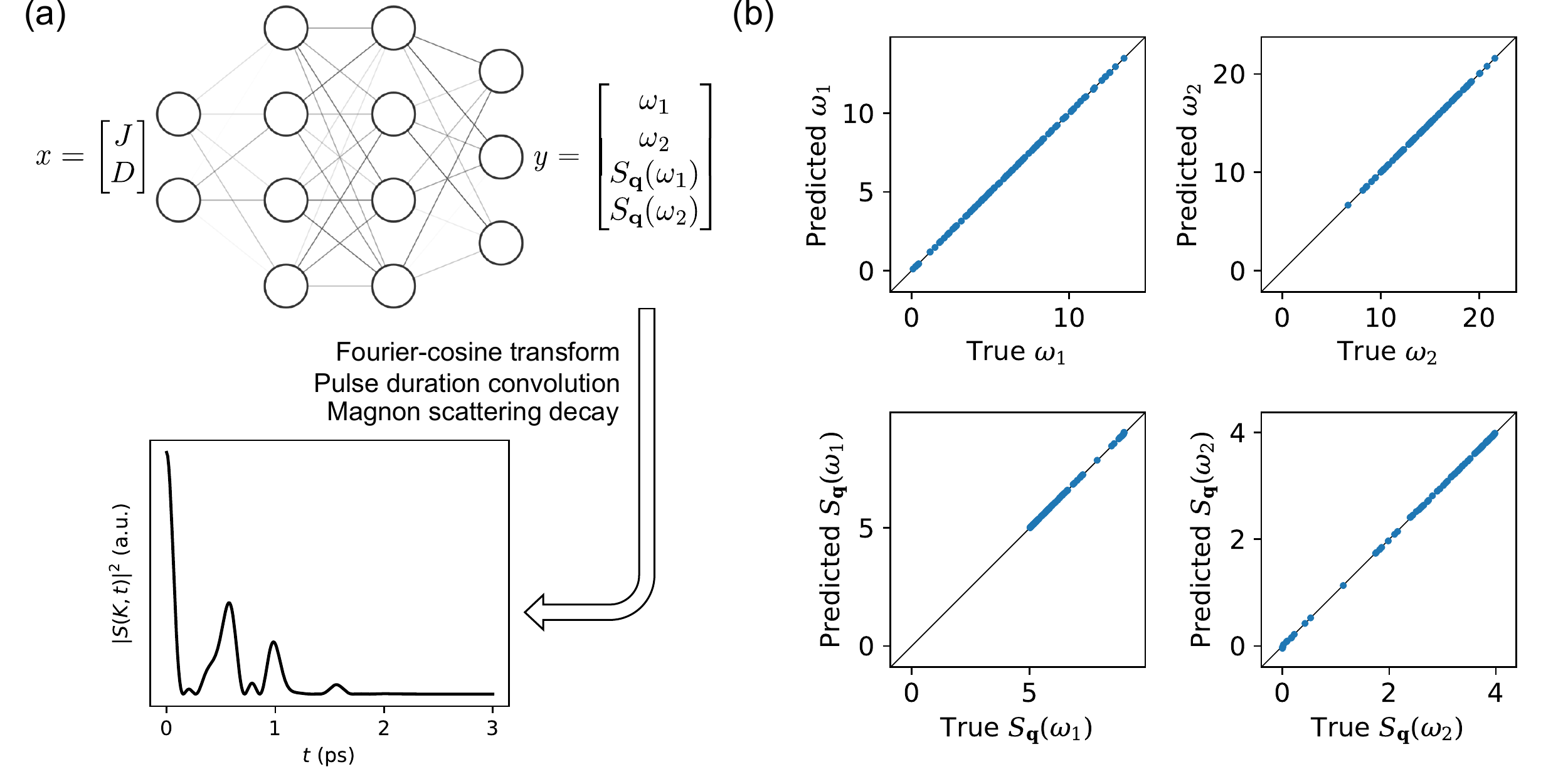}
    \caption{\textbf{Architecture and performance of the machine learning surrogate model.} \textbf{(a)} Neural network-based surrogate model for experimental measurement $s(t)$. The calculation from $y$ to $s(t)$ follows Eq.\@ \eqref{eqn:sqt_mag}. \textbf{(b)} Two-dimensional histogram showing excellent agreements between neural network predictions and ground-truth on the testing dataset.}
    \label{fig:network_surrogate_model}
\end{figure}

The surrogate model for spin excitations is a fully connected neural network (NN). As shown in Figure \ref{fig:network_surrogate_model}(a), the network model takes in Hamiltonian parameters $x=[J,D]$ as input and predicts the dynamical structure factor $S(\mathbf{q},\omega)$ in the form of $y=[\omega_{1},\omega_{2},S(\mathbf{q},\omega_{1}),S(\mathbf{q},\omega_{2})]$ at some fixed momentum vector $\mathbf{q}$. Owing to the specific symmetry of the considered Hamiltonian \cite{chen2018topological,zhang2021interplay}, we restrict our attention to dynamics associated with the $K$ reciprocal lattice point (the Dirac point) of $\mathbf{q}=[1/3,1/3,0]$ in this work, which informs the energy gap induced by the DM interaction \cite{owerre2016first,kim2016realization,chen2018topological}. This corresponds to the scenario in an XPFS measurement where the detector is adjusted to capture speckle patterns generated over an area of momentum space which will provide the most valuable information. 

For the preparation of a real experiment, the training dataset, $\{(x_{1},y_{1}),\ldots,(x_{N},y_{N})\}$, needs to be prepared such that it covers a reasonably wide distribution of input parameter space, $x_i=[J_{i},D_{i}]$, to include the unknown ground truth $x^{\ast}$ that describes the measured sample. Here, $J_{i}$ and $D_{i}$ are randomly drawn from uniform distributions, $J_{i}\sim\mathcal{U}(-3, -1)\meV$ and $D_{i}\sim\mathcal{U}(-1, 0)\meV$. In addition, the interlayer exchange interaction and anisotropy parameters are fixed for our demonstration purposes here, such that $J_{\perp}=-0.6\meV$ and $D_{z}=-0.1\meV$. We generate $1,000$ samples in total and randomly allocate $800$ for training, $100$ for validation, and $100$ for testing the NN model. In Figure \ref{fig:network_surrogate_model}(b), we find the trained NN model can predict the magnetic excitations for the testing parameter sets very well.

To obtain quantities $S(\mathbf{q},t)$ that can be extracted from experimental measurements, the network output $y$ is further Fourier-cosine transformed and multiplied by an exponential-decay function with the constant magnon inverse lifetime $\Gamma$. The $\Gamma$ is separated from the network output as the magnon lifetime is not intrinsically connected to the Hamiltonian in the low-temperature limit of LSWT. The resulting function is then convolved with a rectangular function $\Pi(t;a)=h(t+a/2)-h(t-a/2)$ with $h(t)$ being the Heaviside step function, which approximates the effects induced by the finite photon-pulse width $a$ or simulates the instrument resolution, respectively. The mapping from $y$ to $S(\mathbf{q},t)$ is expressed as follows:
\begin{equation}\label{eqn:sqt_mag}
    S(\mathbf{q},t)=\int_{0}^{\infty}e^{-\Gamma \tau}\sum_{i}S(\mathbf{q},\omega_{i})\cos(\omega_{i} \tau) \Pi(t-\tau;a)\ \ud \tau.
\end{equation}
It is worth mentioning that this transformation is differentiable with respect to the inputs, e.g., $\omega_{i}$ and $S(\mathbf{q},\omega_{i})$. Since the direct quantity derived from $g^{(2)}(t)$ is the squared ISF $|S(\mathbf{q},t)/S(\mathbf{q},0)|^{2}$, we denote the $s_{\mathbf{q}}(t)=|S(\mathbf{q},t)/S(\mathbf{q},0)|^{2}$ and further $s(t)\equiv s_{K}(t)$ for brevity. The complete mapping from $x$ to $s(t)$ is displayed in Figure \ref{fig:network_surrogate_model}(a).

\subsection{Bayesian experimental design and parameter estimation}\label{sec:boed}

\begin{figure}[h]
    \centering
    \includegraphics[width=1.0\linewidth]{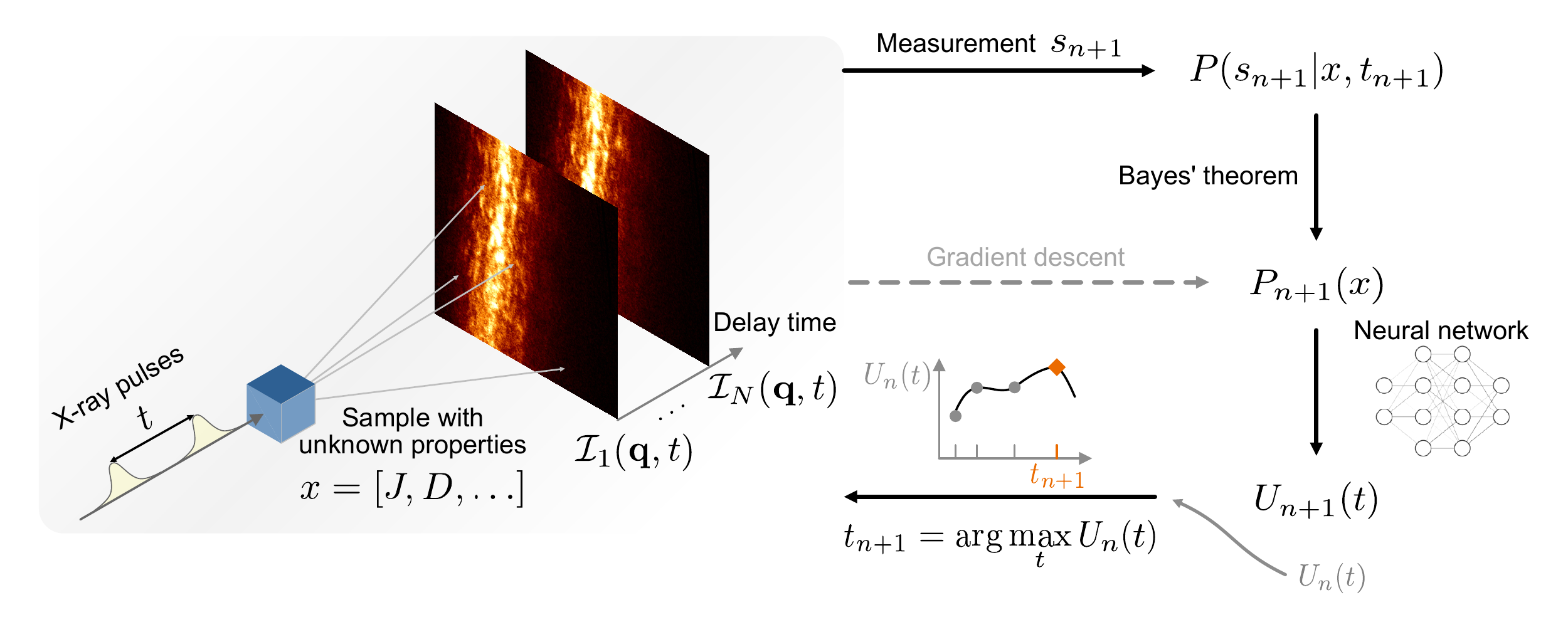}
    \caption{\textbf{A schematic illustration of machine learning-enabled Bayesian experimental design framework.} The speckle pattern shown in the left panel is the magnetic scattering from a van der Waals magnetic system at the LCLS at the Ni $K$-edge, adapted from Ref.\@ \citenum{chen2022testing}.}
    \label{fig:exp_design}
\end{figure}

In this section, we briefly introduce the formulation of BOED in the context of XPFS measurements for magnetic excitations, based on the recent advancements in BOED \cite{granade2012robust, huan2013simulation, ryan2016review, mcmichael2021optbayesexpt, mcmichael2022simplified}. We assume the measurements are taken around the fixed $K$-reciprocal lattice point and at different delay times between two x-ray probe pulses, $t$. The central idea in BOED includes taking full advantage of previously obtained measurement data to make the most informed decisions, and estimating unknown parameters where an analytical model is assumed, as illustrated in Figure \ref{fig:exp_design}. At each step, the distribution of $s(t)$ at each delay time $t$ of the chosen measurement domain $0\le t\le 3 \ps$ is evaluated based on current estimation of the parameter distribution $P(x)$. This distribution of $s(t)$ is further used to calculate the so-called utility function, $U(t)$, that evaluates the ``information gain'' given that the next measurement is taken at $t$, for the entire measurable domain. Upon receiving the new measurement data point, the parameter distribution $P(x)$ can be updated with the Bayes' theorem and the process iterates. We adapted the Python implementation of BOED, \texttt{optbayesexpt}, presented by Refs.\@ \citenum{mcmichael2021optbayesexpt} and \citenum{mcmichael2022simplified} for online parameter learning and experiment steering. We will elaborate each component of the BOED in details below.

Suppose that after measuring $n$ different delay-times, we have the current probability distribution of the model parameters as $P_{n}(x)=P(x|S_{n},T_{n})$, where $S_{n}=\{s_{1},s_{2},\ldots,s_{n}\}$ and $T_{n}=\{t_{1},t_{2},\ldots,t_{n}\}$ are sets of measured data and time points, respectively. The distribution is numerically represented by a group of discrete particles (parameters) $x$ with weights to represent $P_{n}(x)$, which is known as the particle filter method \cite{granade2012robust, elfring2021particle, mcmichael2022simplified}. Here, the utility function $U(t)$ is defined as the $P(s_{n+1}|t_{n+1})$ averaged-distance of distributions between two scenarios: before and after making the measurement $s_{n+1}$ at delay-time $t_{n+1}$. More specifically, it can be expressed through the Kullback–Leibler divergence between $P_{n+1}(x)$ and $P_{n}(x)$, $$D_{\mathrm{KL}}(P_{n+1}(x)||P_{n}(x))=\int P_{n+1}(x) \ln\left[\frac{P_{n+1}(x)}{P_{n}(x)}\right]\ud x,$$ such that 
\begin{equation}\label{eqn:util_kld}
\begin{split}
    U_{n}(t)&=\int P_{n+1}(s|t) D_{\mathrm{KL}}(P_{n+1}(x)||P_{n}(x))\ \ud s\\
    &=\int P_{n}(x)\left[ \int P_{n+1}(s|x,t) \ln P_{n+1}(s|x,t)\ \ud s\right]\ud x  - \int P_{n+1}(s|t) \ln P_{n+1}(s|t)\ \ud s,
\end{split}
\end{equation}
where we used shorthands $P_{n+1}(s|t)$ and $P_{n+1}(s|x,t)$ for $P(s_{n+1}|t_{n+1})$ and $P(s_{n+1}|x,t_{n+1})$. The first term in Eq.\@ \eqref{eqn:util_kld} represents the negative expected differential entropy of noise distributions only, while the second term is the differential entropy of $P_{n+1}(s|t)$, i.e., the distribution of predicted signal at $t$ being $s$ with the current parameter distribution $P_{n}(x)$. In the ideal case where $P_{n+1}(s|x,t)$ and $P_{n+1}(s|t)$ are both Gaussian distributions, the utility function can be reduced to the form $U_{n}(t)=-\tfrac{1}{2}\ln(\sigma_{\eta}^{2})+\tfrac{1}{2}\ln[\sigma_{\eta}^{2}+\sigma_{s}^{2}(t)]$, where $\sigma_{\eta}$ and $\sigma_{s}(t)$ are standard deviations of measurement noise and noise-free simulated signal. However, the second term can no longer be decomposed as the sum of two variances in the case of Poisson measurement noise. In spite of this restriction, considering that $\sigma_{s}(t)$ reflects the parameter uncertainties in the measurable domain and taking into account the computational efficiency, we simply choose $\sigma_{s}^{2}(t)$ as our basic utility function in this work. A detailed discussion about this simplification is provided in Appendix \ref{sec:derivation_util}. In practice, $\sigma_{s}^{2}(t)$ is calculated by passing $x\sim P_{n}(x)$, i.e., the particles (parameters) used to represent $P_{n}(x)$, into the NN surrogate model introduced in Section \ref{sec:ml_model} and evaluating the (particle weights-weighted) variance on the ensemble of outputs.

In addition to the bare variance $\sigma_{s}^{2}(t)$ , we consider a cost function of form $c(t)=1 + h \sum_{n=1}^{N}\exp[-\left(\frac{t-t_{n}}{w}\right)^{2}]$ to keep the proposed measurements reasonably apart from each other, where $t_{n}\in T_{N}$ represents previous measured delay-times, and the height and width parameters are $h=10$ and $w=0.25\ps$. Thus, the final effective utility function used in this work is 
\begin{equation}\label{eqn:util_var}
    U_{n}(t)=\frac{\sigma_{s}^{2}(t;P_{n}(x))}{c(t)},
\end{equation}
which provides a rapid assessment of the key information contained in the original expression \eqref{eqn:util_kld}. After calculating the utility function $U_{n}(t)$, the next suggested delay-time point to measure can be obtained by finding the maximizer, namely,
\begin{equation}\label{eqn:t_next_selection}
    t_{n+1}=\argmax_{t} U_{n}(t).
\end{equation}
Upon collection of the new data point $s_{n+1}$ at $t_{n+1}$, the likelihood $P(s|x,t)$ is then given by a Gaussian distribution 
\begin{equation}\label{eqn:likelihood}
    P(s_{n+1}|x,t_{n+1})=\frac{1}{\sqrt{2\pi}\sigma}\exp\left[-\frac{1}{2}\left(\frac{\left|s_{n+1}-\hat{s}_{n+1}\right|}{\sigma}\right)^{2}\right], \qquad \sigma = \sqrt{\eta \max(s_{n+1},1.0)},
\end{equation}
where $s_{n+1}$ and $\hat{s}_{n+1}$ denote measured and predicted signal values, respectively. The Eq.\@ \eqref{eqn:likelihood} used the Gaussian approximation $\mathcal{N}(\mu=\lambda, \sigma=\sqrt{\lambda})$ for the Poisson distribution $\mathcal{P}(\lambda)$ with mean $\lambda$ to approximate the standard deviation $\sigma$, given the Poisson nature of counting photons on detectors. The minimum value clamping $\max(s_{n+1},1.0)$ is used for numerical stability, while the $\eta$ represents the noise level for accounting the effects of signal normalization and is considered known. Although this approximation relies on a sufficiently large photon rate and loses accuracy when the measured signal value $s_{n+1}$ is small, it tends to assign comparably small likelihoods to all $x$ when $\|s_{n+1}-\hat{s}_{n+1}\|$ is large compared with $\sqrt{s_{n+1}}$ and its negative impacts remain limited. Following the calculation of likelihood $P(s_{n+1}|x,t_{n+1})$, the probability distribution of model parameters $x$ after collecting $n+1$ data points is obtained by applying Bayes' theorem,
\begin{equation}\label{eqn:bayesian_update_Pn}
    P_{n+1}(x) = P(x|s_{n+1}, t_{n+1}, S_{n},T_{n}) = \frac{P(s_{n+1}|x,t_{n+1}) P(x|S_{n},T_{n})}{P(s_{n+1}|t_{n+1})}.
\end{equation}
In practice, the posterior is updated using the product of the prior and likelihood
\begin{equation}
    P_{n+1}(x)\propto P(s_{n+1}|x,t_{n+1}) P_{n}(x),
\end{equation}
which is followed by a normalization step to ensure that $\int P_{n+1}(x)\ \ud x=1$.

\begin{figure}[h]
    \centering
    \includegraphics[width=0.9\linewidth]{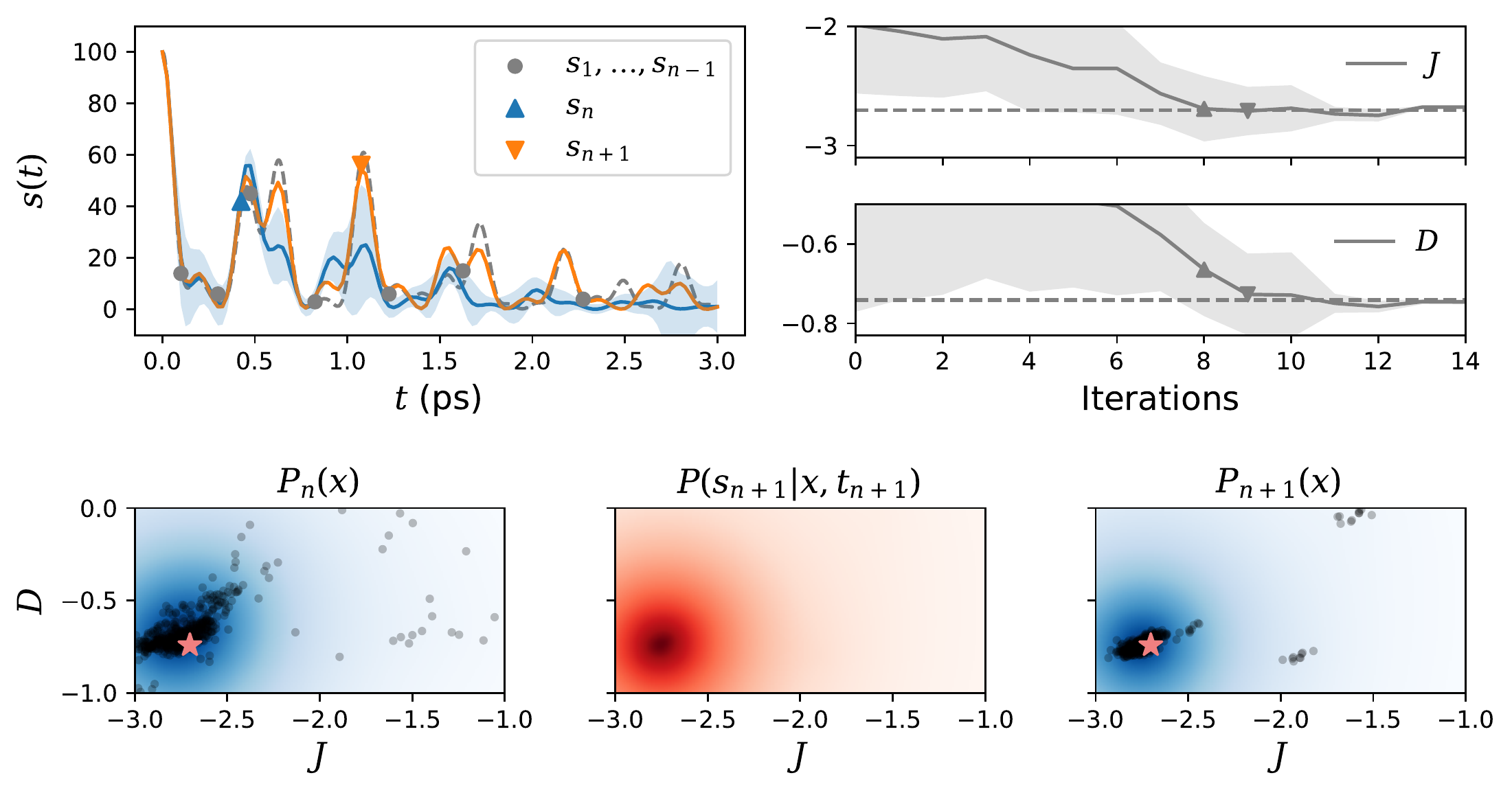}
    \caption{\textbf{Illustrations of Bayes optimal experimental design and parameter estimation.} \textbf{(a)} New measurement is proposed by finding the time delay that maximizes the chosen utility function (the amplitude of blue shaded area). Blue and orange curves represents mean model predictions based on $P_{n}(x)$ and $P_{n+1}(x)$, respectively, for $n=8$. \textbf{(b)} Mean values (solid lines) and standard deviations (shaded area) of key model parameters $J$ and $D$ versus measurement iterations. Joint \textbf{(c)} prior, \textbf{(d)} likelihood, and \textbf{(e)} posterior distributions illustrate the application of Bayes' theorem for the selected iteration given the new measurement $s_{n+1}$ at $t_{n+1}$. The heat-maps (c\textendash e) are obtained by kernel density estimation (KDE) from the original particle representations (particle locations and weights in the particle filter method), while the black markers in (c) and (d) show the corresponding particle locations.}
    \label{fig:bayesian_illustration}
\end{figure}

To better understand the workflow, we illustrate some key elements in Figure \ref{fig:bayesian_illustration}. In Figure \ref{fig:bayesian_illustration}(a), we show the mean model prediction at $n$-th measurement step in the blue curve based on $P_{n}(x)$, with the shaded area representing the magnitude of the utility function $U_{n}(t)$. The suggested time delay for the subsequent measurement is selected to maximize this utility function, as denoted by the orange marker. Following the acquisition of the new data point $s_{n+1}$ at $t_{n+1}$, parameter estimations are updated by applying Bayes' theorem in Eq.\@ \eqref{eqn:bayesian_update_Pn}. The evolving distributions of parameters $J$ and $D$ through the measurement iterations are illustrated in Figure \ref{fig:bayesian_illustration}(b). Additionally, the joint prior, likelihood, and posterior distributions of parameters $x=[J,D]$ are presented in Figures \ref{fig:bayesian_illustration}(c-e). These figures illuminate the step-by-step updating process for the parameter distribution $P_{n}(x)$ by iteratively applying Eqs.\@ \eqref{eqn:util_var}, \eqref{eqn:likelihood}, and \eqref{eqn:bayesian_update_Pn} as shown earlier in Figure \ref{fig:exp_design}.

\subsection{Automatic differentiation-enabled distribution correction}

While BOED is effective in estimating parameter distributions from sequential measurements, it can be limited by chosen prior distributions, making it susceptible to poor initializations \cite{robert2007prior, gelman2013bayesian}. For instance, the posterior distribution will vanish at regions where the prior distribution is assumed to be zero. This can lead to unsuccessful parameter estimation when the true parameters reside outside of the chosen parameter space. However, expanding the parameter space will require a increased number of particles used in the particle filter method to maintain reasonable accuracy in distribution representations. Therefore, it is a non-trivial task to strike a balance between computational efficiency and parameter space size. To address this challenge, we take further advantage of our NN-based forward model. Specifically, we utilize its automatic differentiation (AD) capability to conduct gradient descent (GD) optimizations to update model parameters as indicated by the horizontal dashed gray line in Figure \ref{fig:exp_design}. 

After measuring $N$ delay-time points, one will obtain a measurement dataset $S_{N}$ and a group of prediction datasets $\hat{S}_{N}(x)$ where each containing the model calculations for one of the particles $x$ representing $P_{N}(x)$, i.e., 
$$S_{N}=\{(t_{1},s_{1}),(t_{2},s_{2}),\ldots,(t_{N},s_{N})\}, \qquad \hat{S}_{N}(x)=\{(t_{1},\hat{s}_{1}(x)),(t_{2},\hat{s}_{2}(x)),\ldots,(t_{N},\hat{s}_{N}(x))\}.$$ 
The mean-squared error between measurements and model predictions for each particle $x$ is then calculated as $L(S_{N},\hat{S}_{N})=\frac{1}{N}\sum_{n=1}^{N}(s_{n}-\hat{s}_{n}(x))^{2}$, and subsequently the gradient $\partial L/\partial x$ obtained from AD can be used to update each $x$ through any gradient-based optimization algorithm. 

\begin{figure}[h]
    \centering
    \includegraphics[width=0.8\linewidth]{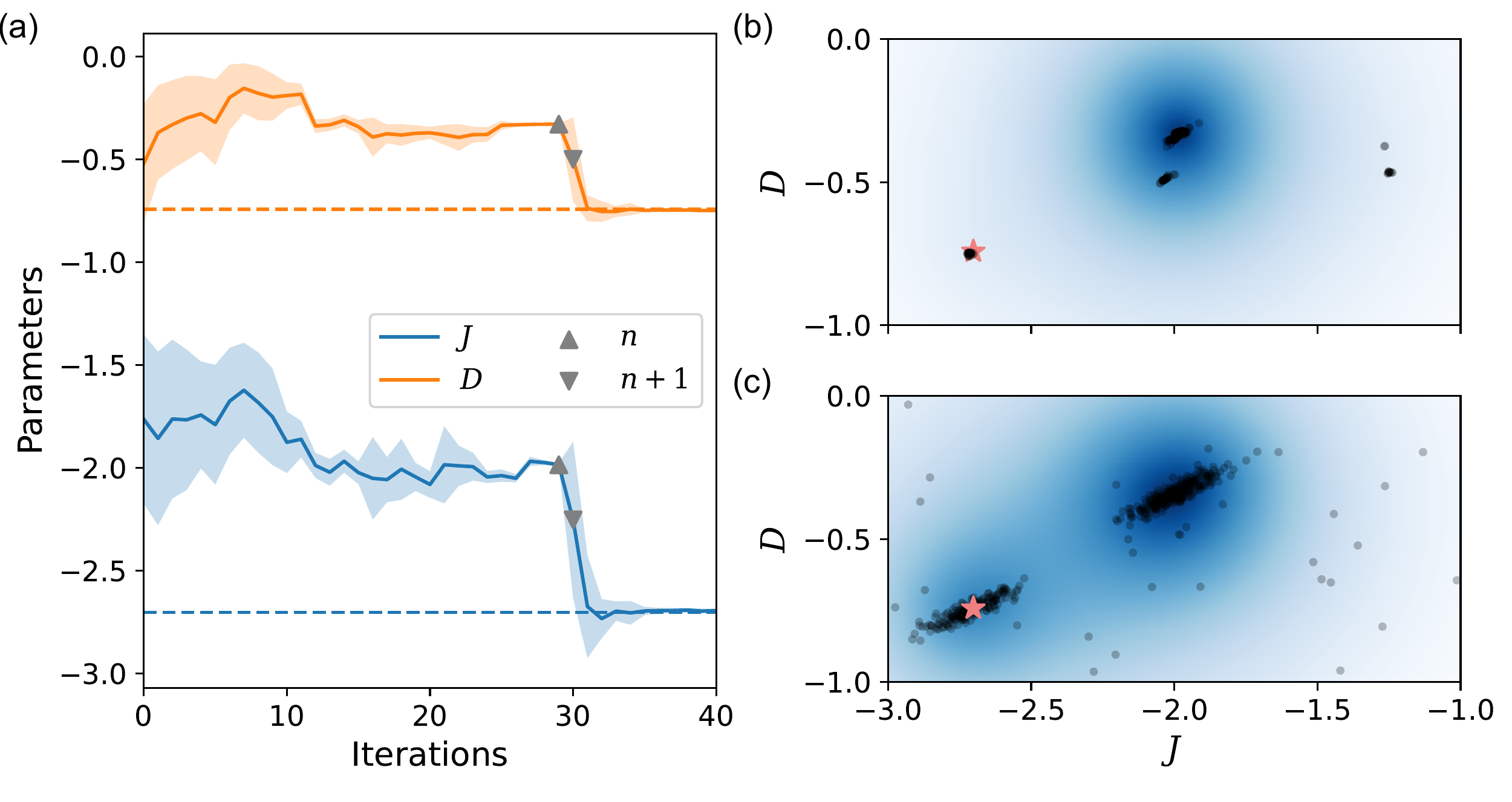}
    \caption{\textbf{Effectiveness of automatic differentiation (AD)-based optimization in correcting false parameter priors.} \textbf{(a)} The pure Bayes method fails to correctly estimate parameter values due to the narrow prior, $J\sim\mathcal{U}(-2.5,-1)$. \textbf{(b)} The resulting parameter distribution has most of its particles concentrated at some values away from the correct value. \textbf{(c)} After application of the AD-based optimization, more particles are brought closer to the true values. In (b--c), the pink star-shaped marker represents the ground-truth $x^{\ast}$.}
    \label{fig:grad_desc_dist}
\end{figure}

In Figure \ref{fig:grad_desc_dist}, we demonstrate the effectiveness of this GD-enhanced BOED strategy for correction of poor priors, where $J$ was initialized from the uniform distribution $J\sim\mathcal{U}(-2.5,-1)$ with its lower bound being higher than the true value $J^{\ast}\approx -2.7$, indicated by the dashed blue line. Initially, the standalone BOED fails to accurately estimate the parameter values correctly, as evidenced by the two plateaus in Figure \ref{fig:grad_desc_dist}(a) preceding the gray markers. The parameter distribution is heavily skewed towards incorrect values, as depicted in Figure \ref{fig:grad_desc_dist}(b). Although the resampling algorithm in the particle filter method succeeds in placing some particles near $x^{\ast}$, their contributions are overwhelmed by the multitude of incorrect estimations. With the application of AD-enabled GD optimization, a greater number of particles are updated to areas close to the true value. While GD optimization does not immediately move all particles near the true values, it effectively overcomes the limitations set by the poor priors and significantly increases the density near $x^{\ast}$, as demonstrated in Figure \ref{fig:grad_desc_dist}(c). This step lays the foundation for BOED to converge to $x^{\ast}$ in subsequent measurement iterations. It is noteworthy that we intentionally choose a smaller parameter space that excludes $x^{\ast}$ for the purpose of demonstration in Figure \ref{fig:grad_desc_dist}. A sensible choice of the initial parameter space can be identical to the parameter space of training dataset, in which case we will show that this hybrid BOED-GD strategy remains effective in achieving better parameter estimations in Section \ref{sec:results_discussion}.

In practice, this GD optimization can be performed intermittently to reach a balance between BOED and GD optimization, e.g., after a certain number of measurement iterations or upon meeting specific criteria. For instance, the GD optimization could be triggered when there are substantial discrepancies between experimental measurements and model predictions and when parameter estimations stagnate.

\section{Results and discussions} \label{sec:results_discussion}

In this section, we present the results on applying our proposed method to simulated experimental data and discuss its performance in comparison with other strategies. We conduct a thorough performance evaluation of four unique experiment steering strategies: random delay-time selection, sequential delay-time selection, and BOED-based selection both with and without GD. These strategies are tested over the same testing dataset containing $100$ samples (different parameters $x_{i}=[J_{i},D_{i}]$) under six distinct experimental conditions, encompassing two x-ray pulse durations, $a=\{0.1,0.2\}\ps$, and three different noise levels, $\eta=\{0.5,1.0,2.0\}$.

\begin{figure}[h]
    \centering
    \includegraphics[width=0.8\linewidth]{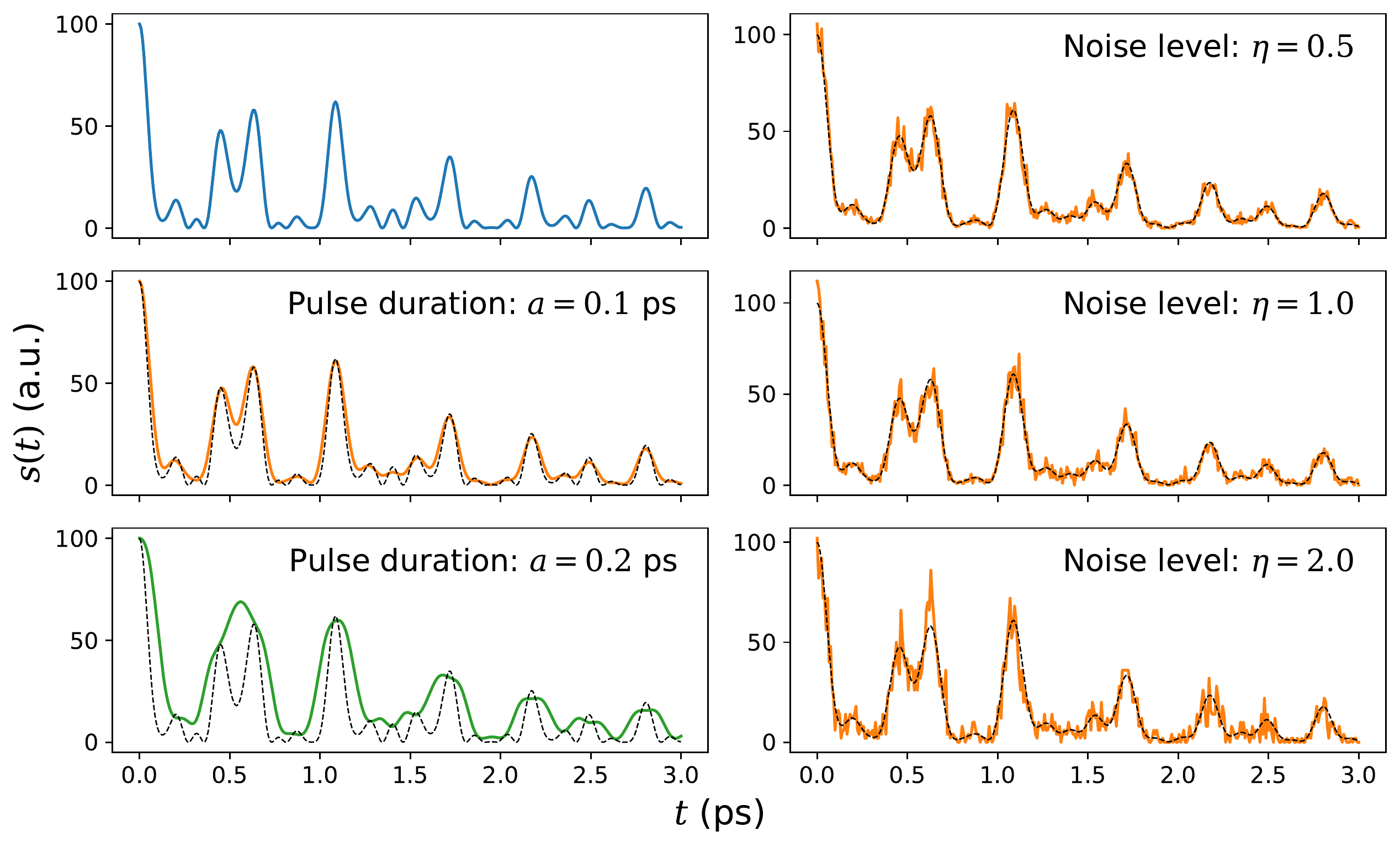}
    \caption{\textbf{Illustration of noise levels and pulse widths.} The top left panel shows the pristine simulated $s(t)$ without noise and pulse duration convolution. The center and lower left panels gives corresponding examples on pulse duration convolutions with $a=0.1$ and $0.2\ps$, respectively, with the black dashed curves being the $s(t)$ shown in the top left panel. The three right panels have different levels of Poisson noise, namely $\eta=0.5$, $1.0$, and $2.0$, on top of the $s(t)$ from the center left panel (the black dashed curves).}
    \label{fig:noise_lvl_pulse_duration}
\end{figure}

To further provide realistic features to the simulations, we add randomly generated Gaussian-shaped lower-energy peaks before calculating $s(t)=|S(K,t)|^{2}$. These extraneous peaks $S^{\mathrm{ext}}(K,\omega)$ are generated based on the highest magnon peak $S_{K}^{\max}=\max\{S(K,\omega):\omega\ge0\}$ by $S^{\mathrm{ext}}(K,\omega)=h_{\mathrm{ext}}\exp[-\omega^{2}/ (2w_{\mathrm{ext}}^{2})]$, where $h_{\mathrm{ext}}$ and $w_{\mathrm{ext}}$ are clamped random variables,
\begin{alignat}{3}
    h_{\mathrm{ext}}&=\max(0,\min(H,S_{K}^{\max})),\qquad &&H& &\sim\mathcal{N}(\mu=S_{K}^{\max}/2, \sigma=S_{K}^{\max}/6), \\
    w_{\mathrm{ext}}&=\max(0.1\meV,\min(W,1.5\meV)),\qquad &&W& &\sim\mathcal{N}(\mu=0.75\meV, \sigma=0.25\meV).
\end{alignat} 
We do want to emphasize that the successful detection of spin excitations with XPFS will require this perturbing factor to have a weak enough or comparable intensity with magnon-related intensities to avoid obscuring the magnon-induced features in $s(t)$. This could be achieved by particular instrument configurations, e.g., incident angles and photon polarization analysis, or looking at certain area with weak elastic scattering signals in the momentum space. The final $s(t)$ is calculated by 
\begin{equation}
    S(\mathbf{q},t)=S^{\mathrm{mag}}(\mathbf{q},t) + \int_{0}^{\infty}\int_{0}^{\infty}S^{\mathrm{ext}}(\mathbf{q},\omega)\cos(\omega \tau) \Pi(t-\tau;a)\ \ud \omega \ \ud \tau
\end{equation}
at $\mathbf{q}=K$, where the first term $S^{\mathrm{mag}}(\mathbf{q},t)$ follows Eq.\@ \eqref{eqn:sqt_mag}. Specifically, we first normalize $s(t)$ such that $s(0)=100$ and then introduce noise by sampling from the Poisson distribution, $\mathcal{P}(\lambda=s(t)/\eta)$, and multiplying the sampled signal by $\eta$. The varying noise levels can be seen as different signal collection durations. For instance, shorter collection durations would yield lower intensities and signal-to-noise ratios (SNR). Given that we have normalized the signal based on $s(0)$, any experiment-dependent variations in SNR can be addressed by adjusting $\eta$ to align with the experimental SNR. Figure \ref{fig:noise_lvl_pulse_duration} showcases some examples of these distinct experimental conditions.

In each test run for a given strategy and condition, we gather data for $40$ delay-time points. This amount of measurements roughly equates to four to five $12$-hour shifts at the LCLS, which is a common allocation for most beam time proposals. All four strategies employ $501$ particles to represent the distributions using the particle filter method. It is noteworthy that in addition to magnon-related parameters $J$, $D$, and $\Gamma$, we also incorporate parameters related to the lower-energy dynamics that is defined by the Gaussian-shaped peak near $\omega=0$ and thereby $x=[J, D, \Gamma, h_{\mathrm{ext}}, w_{\mathrm{ext}}]$. The prior distributions for each parameter of interest are detailed in Table \ref{tab:priors}. For each iteration $n$, the two Bayes-based strategies are applied to suggest the next measurement $t_{n+1}$ based on the maximum of $U_{n}(t)$ defined in Eq.\@ \eqref{eqn:util_var}, where only those $t$ that have not yet been measured are considered, i.e., $t_{n+1}=\argmax_{t}U_{n}(t\in T_{\mathrm{total}}\setminus T_{n})$. In particular, $T_{\mathrm{total}}$ denotes the set of all measurable delay-times that includes $120$ equally-separated measurable delay-times that spans from $0$ to $3$ ps, i.e., $T_{\mathrm{total}}=\{0.025\times \mathtt{n}\}_{\mathtt{n}=0}^{120}$. For the BOED method incorporating GD optimization, the trigger condition for GD optimization is set as $L(S_{n},\hat{S}_{n})\ge 25$ with $n-n_{\text{last}}\ge 14$, indicating a minimum of 14 iterations between two successive optimization steps. Each GD optimization performs 100 updating steps using the Adam optimizer \cite{kingma2014adam}, with a learning rate of $0.1$. Moreover, the sequential strategy sweeps over the $3\ps$ measurable domain (or all elements in $T_{\mathrm{total}}$) in $40$ iterations such that the measured delay-times are separated by $t_{n+1}-t_{n}=0.075\ps$. The random strategy suggests every new measurement point by randomly drawing a sample from the uniform distribution, $t_{n}\sim \mathcal{U}(T_{\mathrm{total}})$.

\begin{table}[h]
\centering
\caption{\textbf{Marginalized prior distributions $P_{0}(x_{n})$ for the parameter $x_{n}$.} The parameter vector $x$ reads $x=[J,D,\ldots,w_{\mathrm{ext}}]$, and $\mathcal{U}(a,b)$ represents a uniform distribution from $a$ to $b$.}
\label{tab:priors}
\begin{tabular}{cccccc}
\toprule
Parameters $x_{n}$ & $J$ & $D$ & $\Gamma$ & $h_{\mathrm{ext}}$ & $w_{\mathrm{ext}}$ \\
\midrule
Marginalized $P_{0}(x_{n})$ & $\mathcal{U}(-3,-1)$ & $\mathcal{U}(-1,0)$ & $\mathcal{U}(0,1)$ & $\mathcal{U}(0,10)$ & $\mathcal{U}(0.1,2.0)$\\
\bottomrule
\end{tabular}
\end{table}

\begin{figure}[h]
    \centering
    \includegraphics[width=\linewidth]{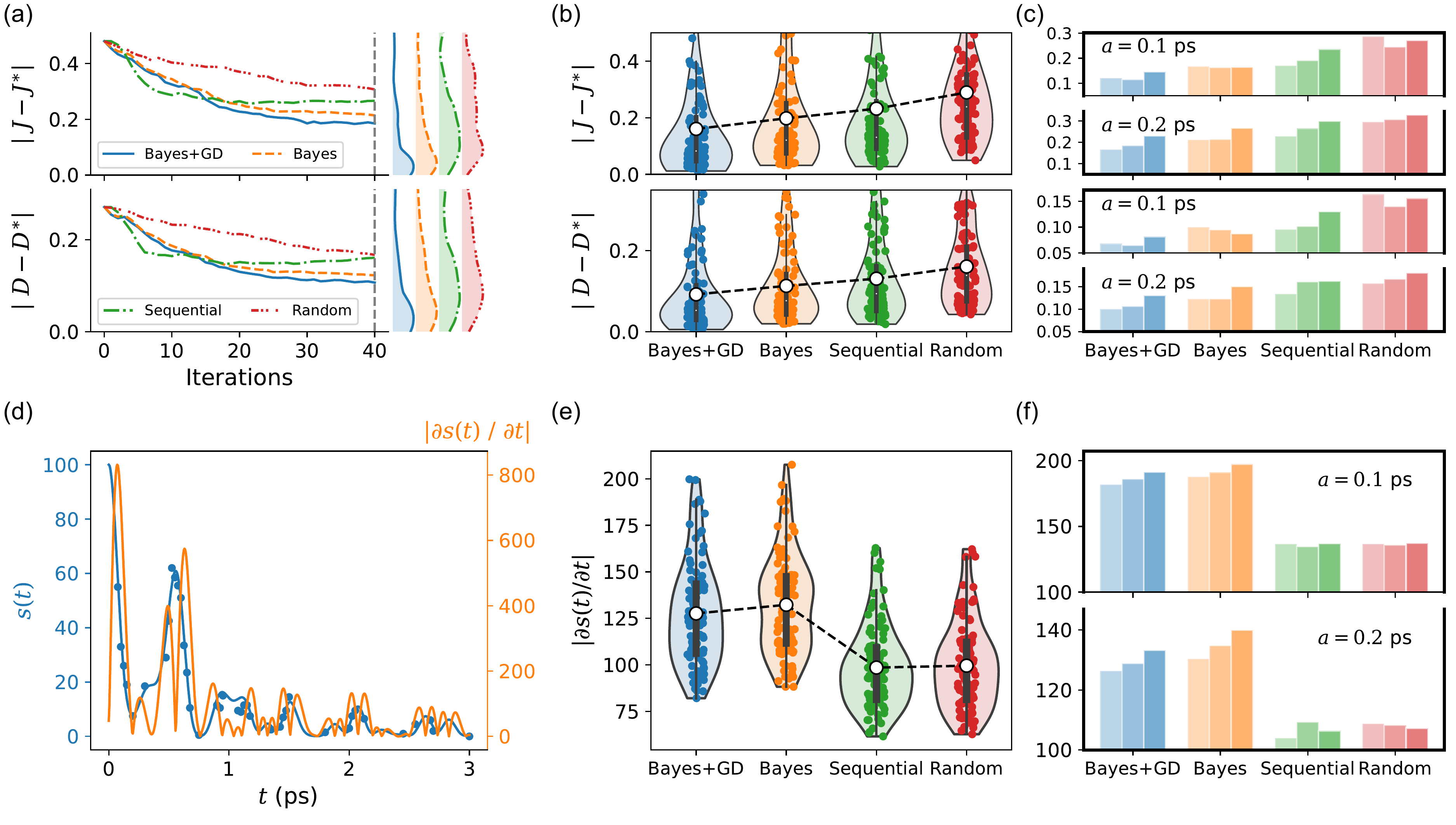}
    \caption{\textbf{A summarized benchmark results over the testing dataset.} \textbf{(a)} Online estimation errors versus measurement iterations for one tested experiment condition ($a=0.1\ps$ and $\eta=1.0$); the right panels illustrate final error distributions at the last measurement iteration. \textbf{(b)} Final error distributions averaged from all six tested conditions $a=\{0.1,0.2\}\ps$ and $\eta=\{0.5,1.0,2.0\}$. \textbf{(c)} Detailed final error distributions for each tested condition. For each of parameters $J$ and $D$, the top and bottom panels display results for $a=0.1$ and $0.2\ps$, respectively. Within each panel, three bars are shown for three noise levels, i.e., bars from left (shallowest color) to right  (darkest color) correspond to $\eta=0.5$, $1.0$, and $2.0$, respectively. \textbf{(d)} A representative example of $|\partial s(t)/\partial t|$ and its original signal $s(t)$. \textbf{(e)} Benchmark of mean $|\partial s(t)/\partial t|$ by four investigated strategies averged from all six tested conditions. \textbf{(f)} Detailed comparison of mean $|\partial s(t)/\partial t|$ for each condition, listed in the same order as in (c).}
    \label{fig:benchmark_summary}
\end{figure}

To obtain more reliable benchmark results for the $4$ different strategies, we conduct five runs for each set of parameters $(J,D)$ under each of the six experimental conditions. In the left panel of Figure \ref{fig:benchmark_summary}(a), we depict the mean absolute errors (MAE) of $J$ and $D$ over $40$ measurement iterations for one specific experimental condition of $a=0.2\ps$ and $\eta=1.0$. For each, the curve strategy is the average of five repeated runs over $100$ testing samples. The distribution of MAE of parameters after the last measurement step is displayed in the right panel. Figure \ref{fig:benchmark_summary}(b) shows the MAE averaged across all six experimental conditions. Here, we observe that the combined BOED-based strategies outperform the the sequential and random measurement strategies. Remarkably, the combined BOED-GD strategy results in the lowest online learning error. The detailed comparisons are presented in Figure \ref{fig:benchmark_summary}(c), where the top two panels provide the MAE after the last measurement iteration for $J$ under two different pulse durations: top and bottom panels for $a=0.1\ps$ and $0.2\ps$, respectively. Each panel depicts the MAE under three different noise levels for each strategy, namely, the left, middle, and right bars for $\eta=0.5$, $1.0$, and $2.0$, respectively. In general, lower online estimation errors are expected for shorter pulse durations and lower noise levels.

In addition to online learning accuracy, the informativeness of the data collected using different strategies is another significant factor to consider. We consider the absolute values of the derivative, $|\partial s(t)/\partial t|$ as an indicator of the informativeness of a measurement. In general, delay-times with higher absolute derivatives carry more information about curve profiles as they cover regions where $s(t)$ changes more rapidly, especially at both sides of extrema of $s(t)$, as illustrated in \ref{fig:benchmark_summary}(d). These measurements with higher $|\partial s(t)/\partial t|$ may lead to better offline parameter fitting results since the extrema reflect physical information about the magnon excitation modes in the time domain. We summarize the comparison of the average informativeness in Figure \ref{fig:benchmark_summary}(e), which is derived from all $40$ recommended delay-times over $100$ testing samples across $6$ experimental conditions. We find that the BOED-based methods significantly outperform the other methods in terms of this informativeness indicator. More detailed comparisons per pulse duration and noise level are presented in Figure \ref{fig:benchmark_summary}(f).

\begin{figure}[h]
    \centering
    \includegraphics[width=\linewidth]{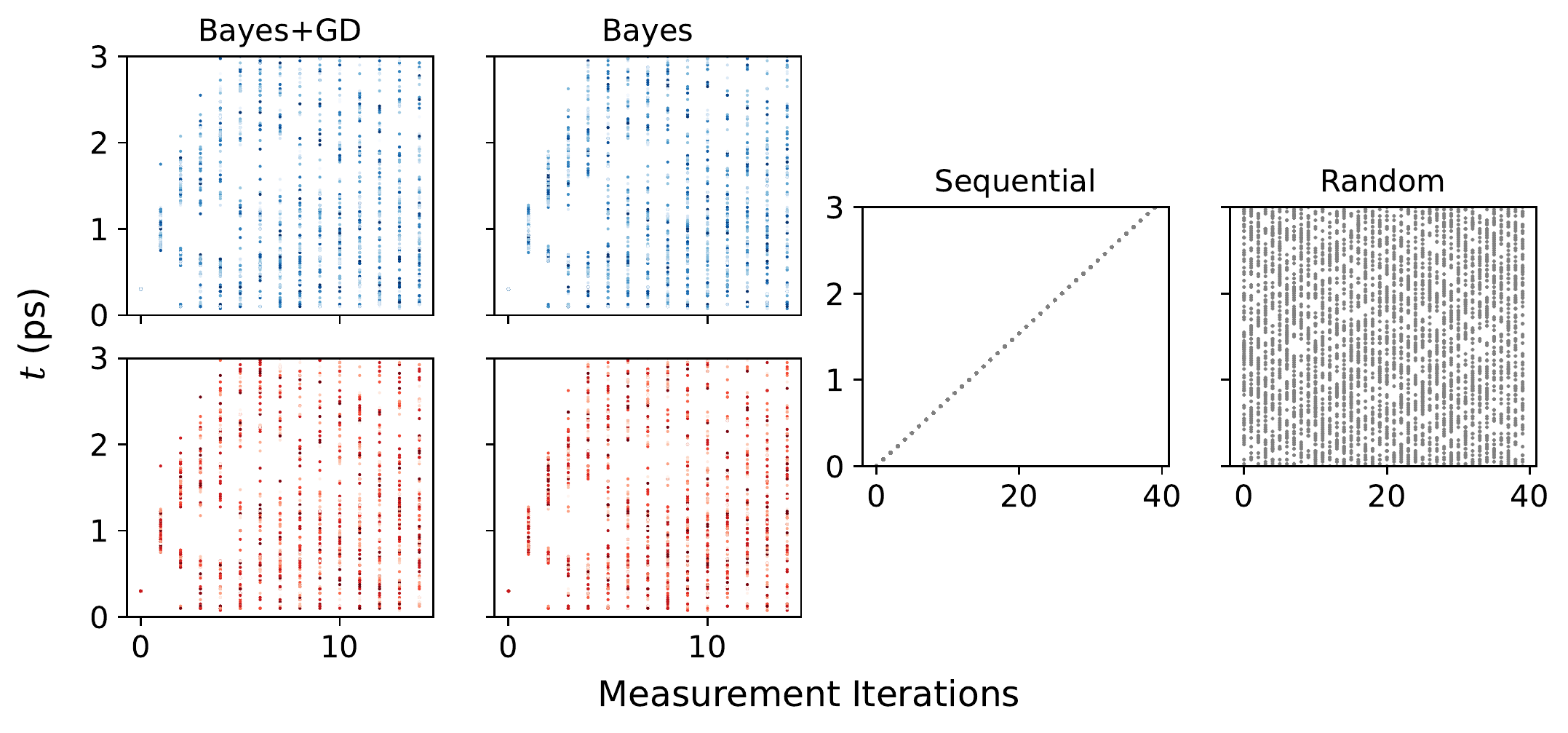}
    \caption{\textbf{Visualization of suggested delay-time measurements by different strategies.} The Bayes-based strategies has two panels where the top panel is colored by values of $J$ and the bottom panel by $D$, the positions of plotted points are same. The BOED-based strategies have the same initial suggestion $t_{0}$ since the same set of randomly sampled prior parameters is used throughout the benchmarking test.}
    \label{fig:settings_visualization_zoom}
\end{figure}

To offer a macroscopic perspective on our benchmark results, we plot the measurements suggested by each strategy from one run (out of their five test runs) across all testing samples as a function of measurement iterations in Figure \ref{fig:settings_visualization_zoom}. The left two panels for BOED-based methods are color-coded in blue and red by the parameters $J$ and $D$, respectively, which have been zoomed into the first $15$ iterations for better visualization and the complete setting visualizations are provided in Appendix \ref{sec:appendix_settings} and Figure \ref{fig:settings_visualization}. It is noticeable that the suggestion patterns of the BOED-based methods initially focus on local domains around $1\ps$ and subsequently branch out to cover the entire measurable domain. This implies that these delay-times around $1\ps$ contain at first the most critical information to narrow down the Hamiltonian parameterizations. In subsequent measurement steps, the BOED-based methods strategically avoid measurements in close proximity to these domains, as demonstrated by the blank areas in Figure \ref{fig:settings_visualization_zoom}, and instead explore broadly throughout both smaller and larger delay-times. Ultimately, the entire measurable domain is filled up as each parameter set necessitates measurements from distinct regions to capture corresponding nuanced features. In contrast, the sequential and random strategies yield either trivially linear or uniformly dispersed patterns as shown in Figure \ref{fig:settings_visualization_zoom}, representing two distinct extremes in experimental measurement design.

The distributions of suggested measurements in the first few iterations can also illuminate where the most distinctive features of different parameter values appear in the time domain. In general, the patterns in the first few iterations displayed by the BOED-based methods in Figure \ref{fig:settings_visualization_zoom} can be interpreted as an estimated significance ranking of the time domains, where a domain appearing in earlier iterations suggests a higher priority for measurements. Thus, in case there is no condition to run the proposed method online during data collection, an alternate application of our method could involve preparing these scatter plots from simulated data prior to the beam time. Then, delay-time measurements can be arranged based on the importance of each time domain informed by the pre-prepared scatter plots.

\section{Concluding remarks} \label{sec:concluding_remarks}

In this work, we presented a ML-enabled BOED method for measuring magnetic excitations with XPFS. We leverage a NN-based forward model to serve as a surrogate model for LSWT in order to enable massive forward computations that are crucial to precise distribution estimations and utility function calculations in BOED. We further incorporate the capability of automatic differentiation from our NN-based forward model in the BOED workflow as a complementary parameter optimization method. A thorough benchmarking conducted over $100$ testing samples under six distinctive experimental conditions demonstrated the superior performance of the ML-enabled BOED methods, with the joint Bayes and GD strategy being especially noteworthy.

Despite these promising results, there is still room for further improvements in terms of broadening application scenarios and refining utility functions. Currently, our method is developed for suggesting delay-time measurements only. More advanced machine learning techniques can be adopted into the presented framework to predict more complicated spin excitations and guide more ``tuning knobs'' (experimental parameters) in experiments. For instance, while the current work focuses on a single momentum vector $\mathbf{q}=K$ with the NN model predicting a vector including $S(K,\omega)$ for a fixed number of energies $\omega$, the proposed method can be readily incorporated with alternative NN surrogate models. For instance, one could employ implicit neural representation-based forward models for $S(\mathbf{q},\omega)$ that takes in continuous momenta $\mathbf{q}$ and energies $\omega$ as inputs and outputs a scalar of the $S(\mathbf{q},\omega)$ as demonstrated in Ref.\@ \citenum{chitturi2023capturing}, which could be utilized to capture more excitation modes and guide measurements in the momentum space. Moreover, the effective utility function is considerably simplified from the original KL divergence-based expression and may not be a faithful representation for the full expression. Future work could explore policy optimization and reinforcement learning techniques for the development of faster and more powerful NN-based utility functions. An interesting extension of our method would be considering additional interactions in Hamiltonian such as the Kitaev interaction, which provides us a theoretical platform to perform estimations on competing parameters between nearly indistinguishable dispersion features \cite{lee2020fundamental, zhang2021interplay}. 

In summary, this proposed framework combines conventional BOED methods and ML techniques in a synergistic manner. It allows for more informed experimental planning, harnessing the power of physical models and Bayes' theorem to collect more meaningful data and richer information within same amount of allocated beam time. The generality of the network modeling allows it to be trained with simulation data generated by more advanced computational methods like exact diagonalization (ED) and density matrix renormalization group (DMRG) to can capture other fundamental spin excitations beyond descriptions of LSWT. Moreover, this method can be applied to guide measurements beyond magnetic systems and XPFS measurements, such as in guiding time-resolved resonant inelastic x-ray scattering with a surrogate model for ED \cite{chen2019theory, mitrano2020probing}. The increased information gain provided by this method will facilitate more efficient measurements and expedite the capture of complex physical phenomena. We expect the developed method to greatly benefit simulation-based experiment planning and eventually accelerate scientific discovery.

\section*{Acknowledgement}

This work was supported by the U.S. Department of Energy, Office of Science, Basic Energy Sciences under Award No.\@ DE-SC0022216. Portions of this work were also supported by the U.S.\@ Department of Energy, Office of Science, Basic Energy Sciences, through the Materials Sciences and Engineering Division, as well as the Scientific User Facilities Division through the Linac Coherent Light Source (LCLS), SLAC National Accelerator Laboratory, both operating under under Contract DE-AC02-76SF00515. J.\@ J.\@ Turner acknowledges support from the U.S.\@ DOE, Office of Science, Basic Energy Sciences through the Early Career Research Program. C.\@ Peng thanks Lichuan Zhang for the enlightening discussion. Z.\@ Chen is grateful for the insightful comments and suggestions from Dr.\@ Robert D.\@ McMichael, and acknowledges the assistance of the large language model ChatGPT by OpenAI in refining the language and enhancing the readability of this paper.

\bibliography{references}


\renewcommand{\thefigure}{A\arabic{figure}}
\setcounter{figure}{0}

\renewcommand{\theequation}{A\arabic{equation}}
\setcounter{equation}{0}
\appendix

\section{Utility function: derivation and simplification} \label{sec:derivation_util}

In this section, we provide more detailed derivation for the utility function as mentioned in Section \ref{sec:boed}. We also discuss the assumptions and simplifications that are adopted towards the final expression \eqref{eqn:util_var}, since we have used the Poisson distribution for measurement noise, which is different from the Gaussian noise assumed in Ref.\@ \citenum{mcmichael2022simplified}. The utility function is constructed as the $P_{n+1}(s|t)$ averaged KL divergence between $P_{n+1}(x)$ and $P_{n}(x)$
\begin{equation}\label{eqn:detail_derivation_Ut}
\begin{split}
    U_{n}(t)&\stackrel{\phantom{\text{Eq.\@ \eqref{eqn:bayesian_update_Pn}}}}{=}\int P_{n+1}(s|t)\ D_{\mathrm{KL}}(P_{n+1}(x)||P_{n}(x))\ \ud s \\
    &\stackrel{\phantom{\text{Eq.\@ \eqref{eqn:bayesian_update_Pn}}}}{=}\int P_{n+1}(s|t) \int P_{n+1}(x) \ln\left[\frac{P_{n+1}(x)}{P_{n}(x)}\right]\ud x \ \ud s \\
    &\stackrel{\text{Eq.\@ \eqref{eqn:bayesian_update_Pn}}}{=}\int P_{n+1}(s|t) \int P_{n+1}(x) \ln \left[\frac{P_{n+1}(s|x,t)}{P_{n+1}(s|t)}\right]\ud x \ \ud s  \\
    &\stackrel{\text{Eq.\@ \eqref{eqn:bayesian_update_Pn}}}{=}\int P_{n+1}(s|t) \int \frac{P_{n}(x)P_{n+1}(s|x,t)}{P_{n+1}(s|t)} \ln P_{n+1}(s|x,t)\ \ud x \ \ud s - \int P_{n+1}(s|t) \ln P_{n+1}(s|t)\ \ud s \\
    &\stackrel{\phantom{\text{Eq.\@ \eqref{eqn:bayesian_update_Pn}}}}{=}\int P_{n}(x)\left[ \int P_{n+1}(s|x,t) \ln P_{n+1}(s|x,t)\ \ud s\right]\ud x  - \int P_{n+1}(s|t) \ln P_{n+1}(s|t)\ \ud s,
\end{split}
\end{equation}
which gives us the Eq.\@ \eqref{eqn:util_kld}. Recalling the definition of differential entropy
\begin{equation}
    H[P(x)]=-\int P(x)\ln P(x)\ \mathrm{d}x,
\end{equation}
it can be found by inspecting the last row of Eq.\@ \eqref{eqn:detail_derivation_Ut} that
\begin{equation}
    U_{n}(t)=-\int P_{n}(x)\ H[P_{n+1}(s|x,t)]\ \ud s\ \ud x  + H[P_{n+1}(s|t)].
\end{equation}
In particular, $P_{n+1}(s|x,t)$ in the first term basically represents measurement noise distribution when the delay-time $t$ and model parameter $x$ are both fixed. 

Suppose $P_{n+1}(s|x,t)$ is Gaussian distribution, or Gaussian-approximated Poisson distribution, with certain variance $\sigma_{\eta}^{2}$, the first term is simply $\mathbb{E}_{x}[-\tfrac{1}{2}\ln \sigma_{\eta}^{2}]+\mathrm{const}$. The dependence on $x$ comes from the fact that the Gaussian approximation for Poisson distribution has a varying standard deviation depending on the calculated signal value $\sigma_{\eta}(t;x)\sim \sqrt{s(t;x)}$. However, when the noise is Gaussian with fixed variance $\sigma_{\eta}^{2}$, the first term reduces to $-\tfrac{1}{2}\ln \sigma_{\eta}^{2}+\mathrm{const}$. 

\begin{figure}[h]
    \centering
    \includegraphics[width=0.6\linewidth]{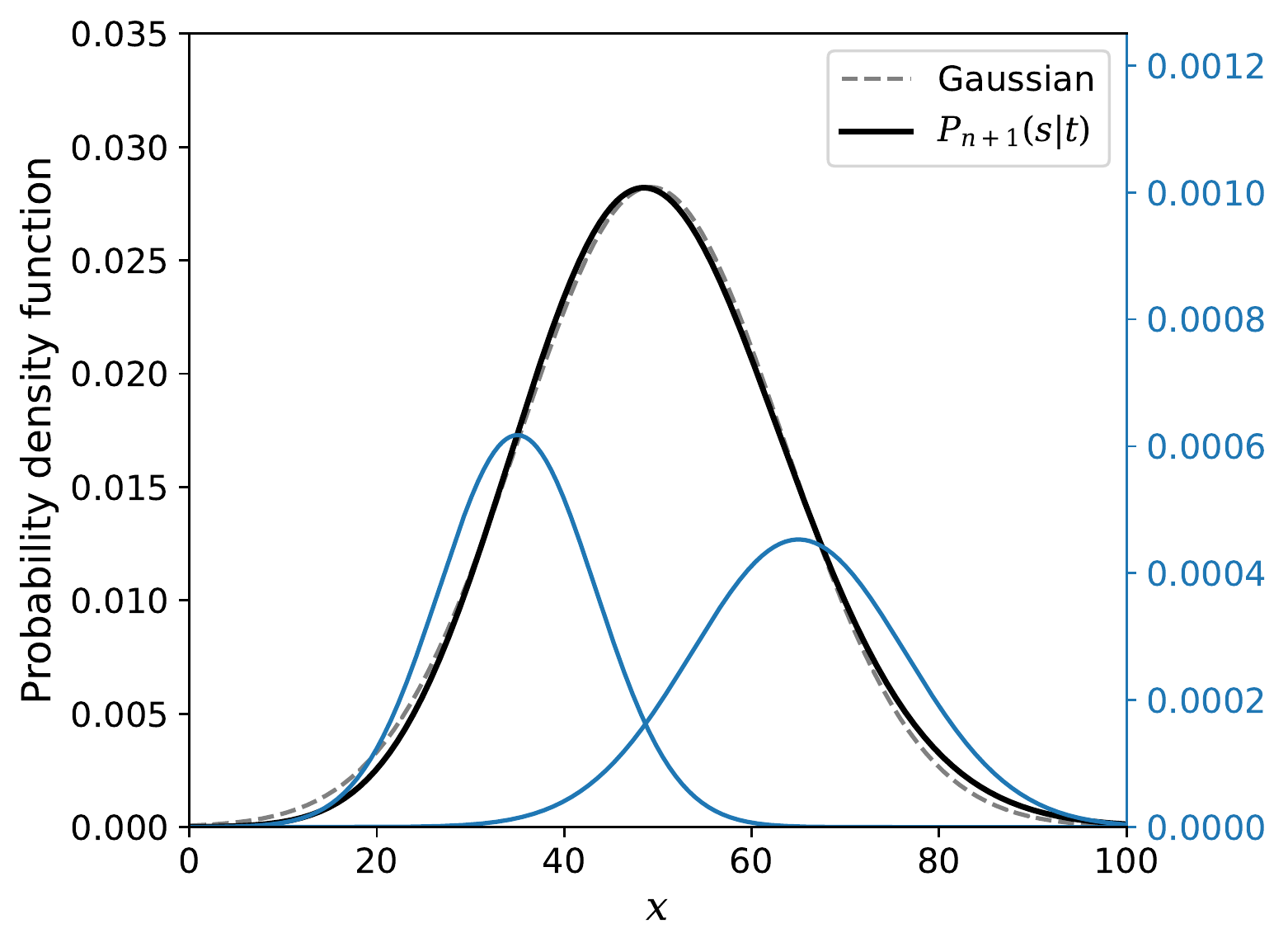}
    \caption{\textbf{Comparison between a representative $P_{n+1}(s|t)$ and a fitted Gaussian probability density.} We used $P_{n+1}^{\mathrm{sim}}(s|t)=\mathcal{N}(\mu=50,\sigma=10)$ as the distribution of noise-free simulated signals, and the noise distribution is simply $\mathcal{N}(s, \sqrt{\eta s})$ with $\eta=2.0$. Two representative noise distributions centered at $s=35$ and $65$ are shown in blue curved. The black solid curve represents the final convolution results for $P_{n+1}(s|t)$, which can be well approximated by a Gaussian $\mathcal{N}(\mu=49.25, \sigma=14.13)$, shown in the dashed gray curve.}
    \label{fig:P_s_given_t}
\end{figure}

If the noise is Gaussian with fixed variance $\sigma_{\eta}^{2}$ and when the calculated noise-free $s(t;x)$ from parameter distribution $P(x)$ is also Gaussian with variance $\sigma_{s}^{2}$ (say, $P_{n+1}^{\mathrm{sim}}(s|t)\sim\mathcal{N}(\mu_{s},\sigma_{s})$) , the second entropy becomes $H[P_{n+1}(s|t)]=\tfrac{1}{2}\ln (\sigma_{\eta}^{2}+\sigma_{s}^{2})+\mathrm{const}$, owing to the fact that $P_{n+1}(s|t)$ is the convolution of two Gaussian distributions. These assumptions leads to the expression $U_{n}(t)=\ln(1+\sigma_{s}^{2}/\sigma_{\eta}^{2})$. Strictly speaking, this decomposition no longer holds when Poisson noise is adopted, since now the noise distribution has signal-dependent variances. This can be demonstrated by the two blue curves in Figure \ref{fig:P_s_given_t}, which represents Gaussian approximated noise distributions when $s=35$ and $65$, respectively. However, we numerically show that $P_{n+1}(s|t)$ can still be well approximated by a Gaussian distribution, as illustrated by comparing the black curve and the dashed gray curve in Figure \ref{fig:P_s_given_t}. This implies that we can still safely approximate $H[P_{n+1}(s|t)]$ with $\tfrac{1}{2}\ln \sigma_{\mathrm{total}}^{2}+\mathrm{const}$. 

In practice, given that the $\sigma_{s}^{2}$ can be directly calculated from $P_{n}(x)$, while $\sigma_{\mathrm{total}}^{2}$ requires additional calculations, and considering that the utility is monotonically increasing function of $\sigma_{s}^{2}$, we simply take $\sigma_{s}^{2}$ as our effective utility function. Since the noise variance $\sigma_{\eta}^{2}(t)$ is no longer constant across the measurement domain, this simplification could lead to different suggested delay-times from those calculated from the original utility function. However, $\sigma_{s}^{2}$ still provides the key information included in \eqref{eqn:util_kld} and serves as a good effective utility measure as discussed in Section \ref{sec:results_discussion}. Further improvements on the effective utility function are left for future studies.

\clearpage
\section{Full view of settings} \label{sec:appendix_settings}

\begin{figure}[h]
    \centering
    \includegraphics[width=\linewidth]{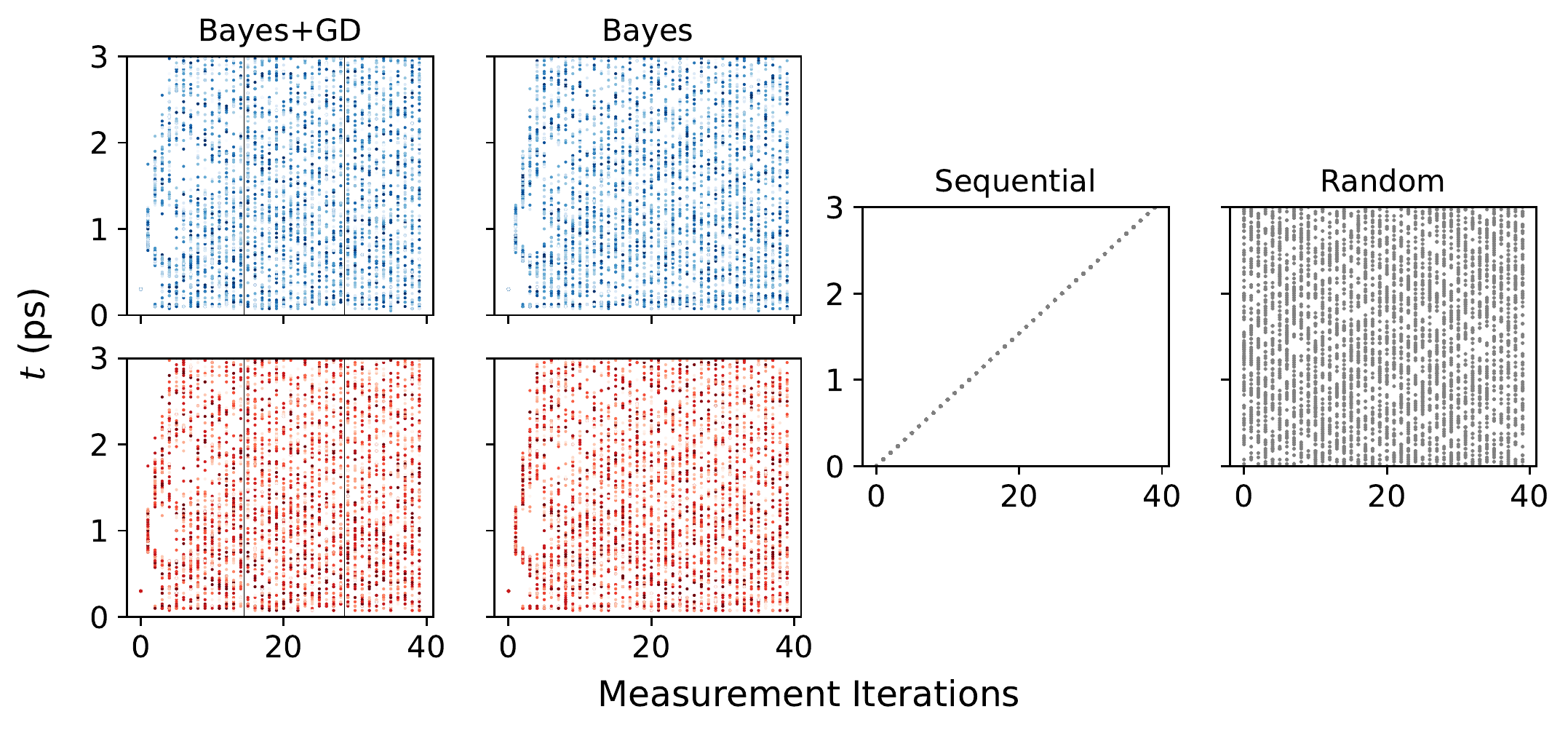}
    \caption{\textbf{Visualization of suggested delay-time measurements by different strategies.} The Bayes-based strategies has two panels where the top panel is colored by values of $J$ and the bottom panel by $D$, the positions of plotted points are same. This figure displays full settings for the BOED-based strategies.}
    \label{fig:settings_visualization}
\end{figure}

We present the full settings suggested by two BOED-based strategies in Figure \ref{fig:settings_visualization}. Different from Figure \ref{fig:settings_visualization_zoom} in the main text, there is no cutoff in the $x$-axis for the left two panels.

\end{document}